# Experimental evidence for direct insulator-quantum Hall transition in multi-layer graphene

Chiashain Chuang[1,2], Li-Hung Lin[3], Nobuyuki Aoki[2], Takahiro Ouchi[2], Akram M. Mahjoub[2], Tak-Pong Woo[1], J. P. Bird[2,4], Yuichi Ochiai[2], Shun-Tsung Lo[5], and Chi-Te Liang[1,5]

[1]Department of Physics, National Taiwan University, Taipei 106, Taiwan
[2]Graduate School of Advanced Integration Science, Chiba University, Chiba 263-8522, Japan
[3]Department of Electrophysics, National Chiayi University, Chiayi 600, Taiwan
[4]Department of Electrical Engineering, University at Buffalo, The State University of New York, Buffalo, New York 14206-1500, USA
[5]Graduate Institute of Applied Physics, National Taiwan University, Taipei 106, Taiwan
*Corresponding author: n-aoki@faculty.chiba-u.jp; ctliang@phys.ntu.edu.tw

We have performed magnetotransport measurements on a multi-layer graphene flake. At the crossing magnetic field $B_c$, an approximately temperature-independent point in the measured longitudinal resistivity $\rho_{xx}$, which is ascribed to the direct insulator-quantum Hall (I-QH) transition, is observed. By analyzing the amplitudes of the magnetoresistivity oscillations, we are able to measure the quantum mobility $\mu_q$ of our device. It is found that at the direct I-QH transition, $\mu_q B_c \approx 0.37$ which is considerably smaller than 1. In contrast, at $B_c$, $\rho_{xx}$ is close to the Hall resistivity $\rho_{xy}$, i.e., the classical mobility $\mu B_c$ is $\approx 1$. Therefore our results suggest that different mobilities need to be introduced for the direct I-QH transition observed in multi-layered graphene. Combined with existing experimental results obtained in various material systems, our data obtained on graphene suggest that the direct I-QH transition is a universal effect in 2D.

**Introduction**

Graphene, which is an ideal two-dimensional system [1], has attracted a great deal of world-wide interest. Interesting effects such as Berry's phase [2-3] and fractional quantum Hall effect [4-6] have been observed in mechanically-exfoliated graphene flakes [1]. In additional to its extraordinary electrical properties, graphene possesses great mechanical [7], optical [8], and thermal [9] characteristics.

The insulator-quantum Hall (I-QH) transition [10-13] is a fascinating physical phenomenon in the field of two-dimensional (2D) physics. In particular, a direct transition from an insulator to a high Landau-level filling factor $v > 2$ QH state which is normally dubbed as the direct I-QH transition continues to attract interest [14]. The direct I-QH transition has been observed in various systems such as SiGe hole gas [14], GaAs multiple quantum well devices [15], GaAs two-dimensional electron gases (2DEGs) containing InAs quantum dots [16-18], a delta-doped GaAs quantum well with additional modulation doping [19-20], GaN-based 2DEGs grown on sapphire [21] and on Si [22], InAs-based 2DEGs [23], and even some conventional GaAs-based 2DEGs [24], suggesting that it is a universal effect. Although some quantum phase transitions, such as plateau-plateau transitions [25] and metal to insulator transitions [26-29], have observed in single layer graphene, and insulating behaviour has been observed in disordered graphene such as hydrogenated graphene [30-33], graphene exposed to ozone [34], reduced graphene oxide [35] and fluorinated graphene [36-37], the direct I-QH transition has not been observed in graphene-based system. It is worth mentioning that Anderson localization effect, important signature of strong localization which may be affected by a magnetic field applied perpendicular to the graphene plane, was observed in a double layer graphene heterostructure [38], but not in single layer pristine graphene. Moreover, the disorder of single graphene is normally lower than those of multi-layer graphene devices. Since one needs sufficient disorder in order to see the I-QH transition [11], multi-layer graphene seems to be a suitable choice for studying such a transition in a pristine graphene-based system. Besides, the top and bottom layers may isolate the environmental impurities [39-42], making multi-layer graphene a stable and suitable system for observing the I-QH transition.

In this paper, we report magneto transport measurements on a multi-layer graphene flake. We observe an approximately temperature independent point in the measured longitudinal resistivity $\rho_{xx}$ which can be asciibed to experimental evidence for the direct I-QH transition. At the crossing field $B_c$ in which $\rho_{xx}$ is approximately $T$ independent, $\rho_{xx}$ is close to $\rho_{xy}$. In contrast, the product of the quantum mobility determined from the oscillations in $\rho_{xx}$ and $B_c$ is $\approx 0.37$ which is considerably smaller than 1. Thus our experimental results suggest that different mobilities need to be introduced when considering the direct I-QH transition in graphene-based devices.

**Experimental details**

A multi-layer graphene flake, mechanically exfoliated from natural graphite, were deposited onto a 300-nm-thick $SiO_2$/Si substrate. An optical microscopy was used to locate graphene flakes, and the thickness of multi-layer graphene is 3.5 nm, checked by atomic force microscopy (AFM). Therefore the layer number of our graphene device is around ten according to the 3.4 Å graphene inter-layer distance [1, 43]. Ti/Au contacts were deposited on the multi-layer graphene flake by electron-beam lithography and lift-off process. The multi-layer graphene flake was made into a Hall bar pattern with a length-to-width ratio of 2.5 as shown in inset (a) of figure 1 by oxygen plasma etching process [44]. Similar to the work done by disordered graphene, our graphene flakes did not underwent a post exfoliation annealing treatment [45-46]. The magnetoresistivity of the graphene device was measured using standard AC lock-in technique at 19 Hz with a constant current $I = 20$ nA in a He$^3$ cryostat equipped with a superconducting magnet.

**Results and discussion**

Figure 1 shows the curves of longitudinal and Hall resistivity $\rho_{xx}(B)$ and $\rho_{xy}(B)$ at $T = 0.28$ K. Features of magnetoresistivity oscllations accompanied by quantum Hall steps are observed at high fields. In order to further study these results, we analyze the positions of the extrema of the magnetoresistivity oscillations in $B$ as well as the heights of the QH steps. Although the steps in the converted Hall conductivity $\rho_{xy}$ are not well quantized in units of $4e^2/h$, they allow us to determine the Landau level filling factor as indicated in the inset (b) of Fig. 1. The carrier density of our device is calculated to be $9.4 \times 10^{16}$ m$^{-2}$ following the procedure described in Ref. [47-48].

We now turn to our main experimental finding. Figure 2 shows the curves of $\rho_{xx}$ $(B)$ and $\rho_{xy}$ $(B)$ as a function of magnetic field at various temperatures $T$. An approximately $T$-independent point in the measured $\rho_{xx}$ at $B_c = 3.1$ T is observed. In the vicinity of $B_c$, for $B < B_c$, the sample behaves as a weak insulator in the sense that $\rho_{xx}$ decreases with increasing $T$. For $B > B_c$, $\rho_{xx}$ increases with increasing $T$, characteristic of a quantum Hall state. At $B_c$, the corresponding Landau level filling factor is about 125 which is much bigger than 1. Therefore we have observed evidence for a direct insulator-quantum Hall transition in our multi-layer graphene. The crossing points for $B > 5.43$ T can be ascribed to approximately $T$-independent points near half filling factors in the conventional Shubnikov-de Haas (SdH) model [17].

By analyzing the amplitudes of the observed SdH oscillations at various magnetic fields and temperatures, we are able to determine the effective mass $m^*$ of our device which is an important physical quantity. The amplitudes of the SdH oscillations $\rho_{xx}$ is given by [49]

$$\Delta\rho_{xx}(B; T) = 4\rho_0 \exp[\frac{-\pi}{\mu qB}] D(B,T), \qquad (1)$$

where $D(B,T) = \frac{4\pi^3 k_B m^* T}{heB} / \sinh\frac{4\pi^3 k_B m^* T}{heB}$, $\rho_0$, $k_B$, $h$, and e are a constant, the Boltzmann constant, Plank's constant and electron charge, respectively. When $\frac{4\pi^3 k_B m^* T}{heB} > 1$, we have

$$\ln\frac{\Delta\rho_{xx}(B,T)}{T} = C_1 - \frac{4\pi^3 k_B m^* T}{heB} , \qquad (2)$$

where $C_1$ is a constant. Figure 3 shows the amplitudes of the SdH oscillations at a fixed magnetic field of 5.437 T. We can see that the experimental data can be well fitted to Eq. (2). The measured effective mass ranges from $0.06m_0$ to $0.07m_0$ where $m_0$ is the rest mass of an electron. Interestingly, the measured effective mass is quite close to that in GaAs ($0.067m_0$).

In our system, for the direct I-QH transition, near the crossing field $\rho_{xx}$ is close to $\rho_{xy}$. In this case, the classical Drude mobility is approximately the inverse of the crossing field $1/B_c$. Therefore the onset of Landau quantization is expected to take place near $B_c$ [50]. However, it is noted that Landau quantization should be linked with the quantum mobility, not the classical Drude mobility [19]. In order to further study the observed I-QH transition, we analyze the amplitudes of the magnetoresistivity oscillations versus the inverse of $B$ at various temperatures. As shown in Fig. 4, there is a good linear fit to Eq. (1) which allows us to estimate the quantum mobility to be around 0.12 m$^2$/V/s. Therefore near $\mu_q B_c \approx 0.37$ which is considerably smaller than 1. Our results obtained on multilayered graphene are consistent with those obtained in GaAs-based weakly disordered systems [19, 21].

It has been shown that the elementary neutral excitations in graphene in a high magnetic field are different from those of a standard 2D system [51]. In this case, the particular Landau level quantization in graphene yields linear magnetoplasmon modes. Moreover, instability of magnetoplasmons can be observed in layered graphene structures [52]. Therefore in order to fully understand the observed I-QH transition in our multi-layer graphene sample, magnetoplasmon modes as well as collective phenomena may need to be considered. The spin effect should not be important in our system [53]. At present, it is unclear whether intra- and/or inter graphene layer interactions play an important role in our system. Nevertheless, the fact that the low-field Hall resistivity is nominally $T$ independent suggests that Coulomb interactions do not seem to be dominant in our system.

**Conclusion**

In conclusion, we have presented magnetoresistivity measurements on a multi-layered graphene flake. An approximately temperature-independent point in $\rho_{xx}$ is ascribed to the direct I-QH transition. Near the crossing field $B_c$, $\rho_{xx}$ is close to $\rho_{xy}$, indicating that at $B_c$, the classical mobility is close to $1/B_c$ such that $B_c$ is close to 1. On the other hand, $\mu_q B_c \approx 0.37$ which is much smaller than 1. Therefore different mobilities must be considered for the direct I-QH transition. Together with existing experimental results obtained on various material systems, our new results obtained in a graphene-based system strongly suggest that the direct I-QH transition is a universal effect in 2D.

**Abbreviations**

I-QH, Insulator-Quantum Hall; 2DEGs, Two-Dimensional Electron Gases; AFM, atomic force microscopy; SdH, Shubnikov-de Haas; 2D, Two-Dimensional

**Competing interests**

The authors declare that they have no competing interests.

**Authors contributions**

CC and LHL performed the experiments. CC, TO and AMM fabricated the device. NA, YO and JPB coordinated the project. TPW and STL provided key interpretation of the data. CC and CTL drafted the paper. All the authors read and agree the final version of the paper.

**Acknowledgements**

This work was funded by the National Science Council (NSC), Taiwan (grant no: NSC 99-2911-I-002-126 and NSC 101-2811-M-002-096). CC gratefully acknowledges financial support from Interchange Association, Japan (IAJ) and the NSC, Taiwan for providing a Japan/Taiwan Summer Program student grant.

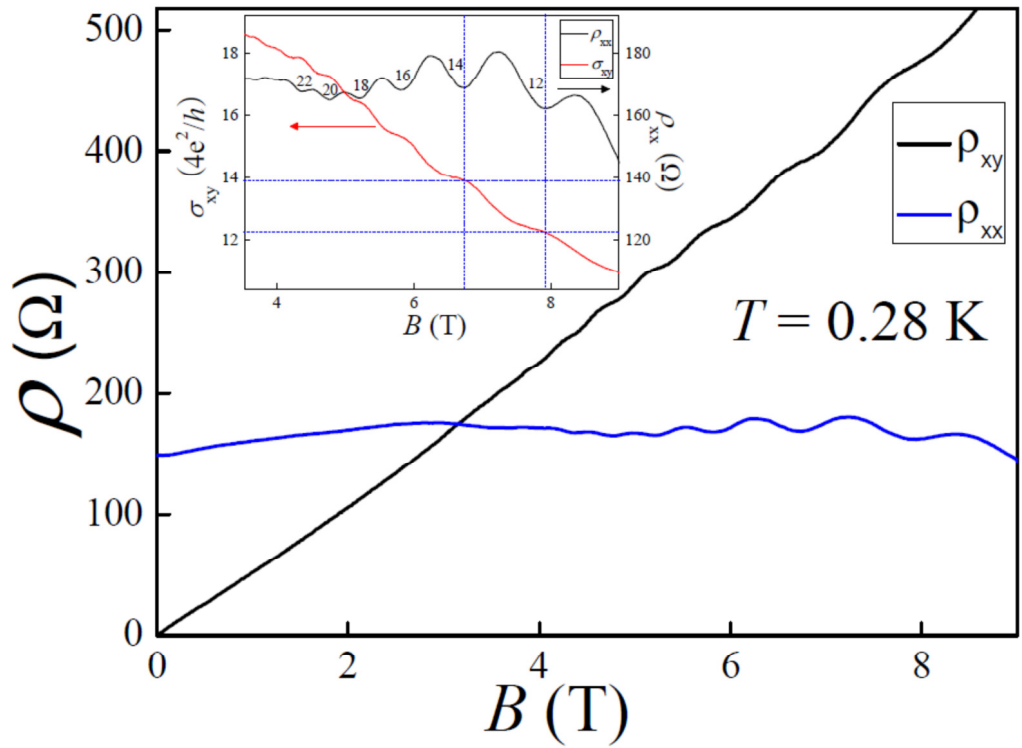

Fig. 1 Longitudinal and Hall resistivity $\rho_{xx}(B)$ and $\rho_{xy}(B)$ as a function of magnetic field at $T = 0.28$ K. Inset (a) shows an optical microscope image of the device. Inset (b) shows the converted $\rho_{xy}$ (in units of $4e^2/h$) and $\rho_{xx}$ as a function of $B$.

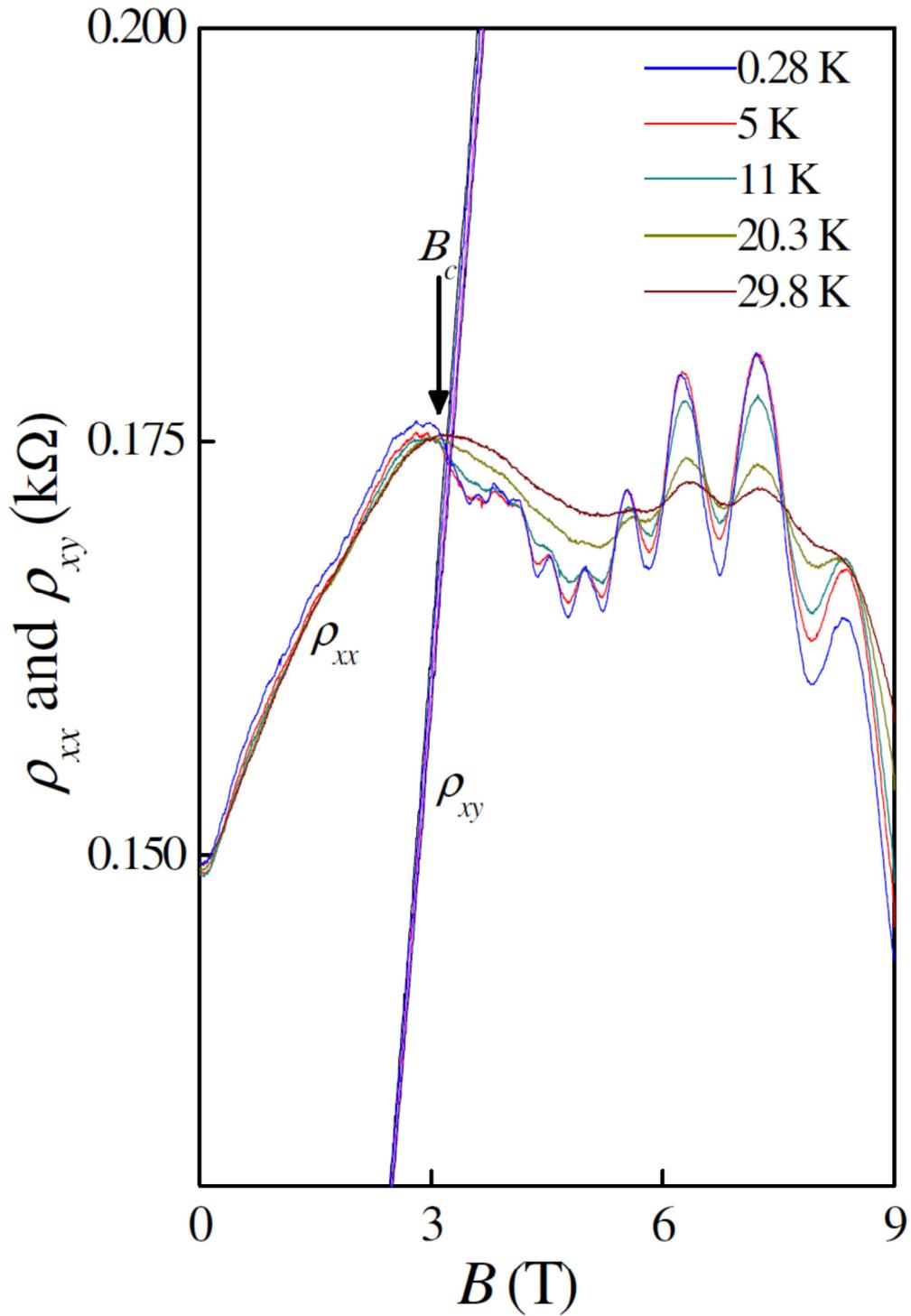

Fig. 2 Longitudinal and Hall resistivity $\rho_{xx}(B)$ and $\rho_{xy}(B)$ as a function of magnetic field at various temperatures $T$. An approximately $T$-independent point in $\rho_{xx}$ is indicated by a crossing field $B_c$.

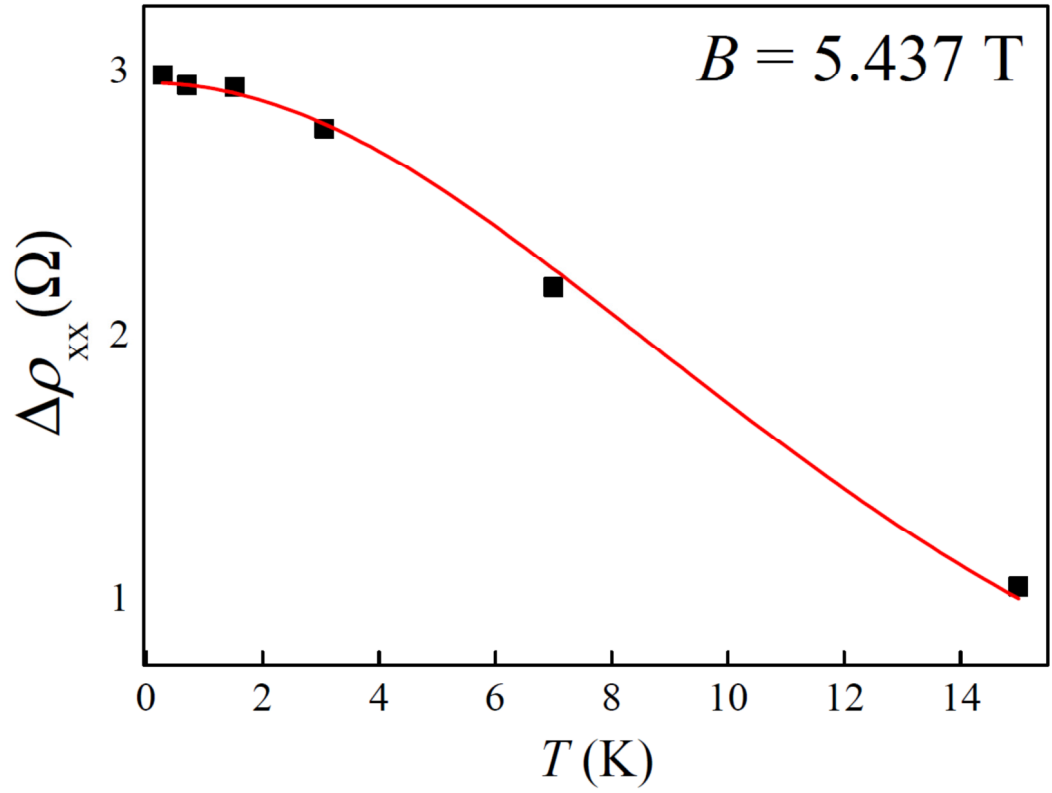

Fig. 3 Amplitudes of the observed oscillations Δρ$_{xx}$ at $B$ = 5.437 T at different temperatures. The curve corresponds to the best fit to Eq. (2).

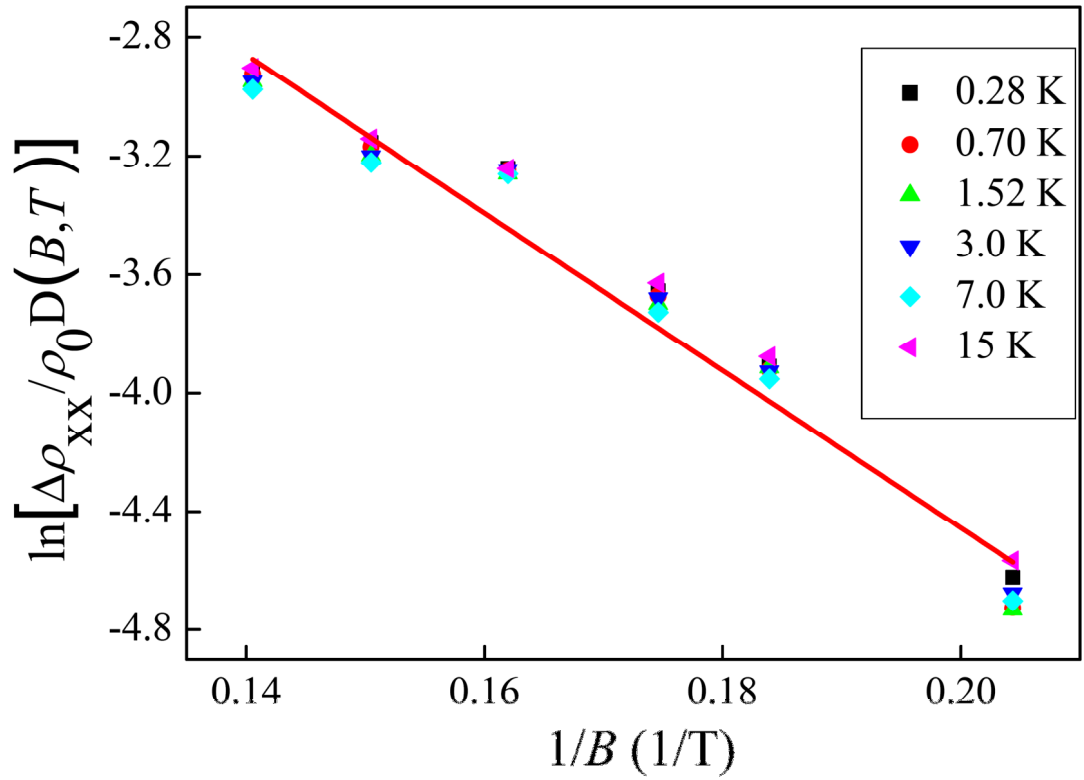

Fig. 4 $\ln[\frac{\Delta\rho_{xx}}{\rho_0 D(B,T)}]$ as a function of the inverse of the magnetic field $1/B$. The solid line corresponds to the best fit to Eq. (1).